\documentclass[twocolumn]{aastex61}
\usepackage{natbib}
\usepackage{amstext}
\usepackage{amsmath}
\usepackage{amssymb}
\usepackage{apjfonts}

\received{November 19, 2017}
\revised{January 23, 2018}
\accepted{January 23, 2018}
\submitjournal{ApJ}

\newcommand{\msun}{M_\odot}

\newcommand{\cc}{{\rm cm^{-3}}}
\newcommand{\nh}{n_{\rm H}}
\newcommand{\nhcen}{n_{\rm H,cen}}
\newcommand{\kms}{{\rm km\,s^{-1}}}
\newcommand{\cs}{c_{\rm s}}

\shorttitle{Formation of first star clusters}
\shortauthors{Hirano et al.}

\begin{document}

\title{Formation of the first star cluster{\bf s} and massive star binaries\\ by fragmentation of filamentary primordial gas clouds}

\correspondingauthor{Shingo Hirano}
\email{shirano@astro.as.utexas.edu}

\author{Shingo Hirano}
\affil{Department of Astronomy, University of Texas at Austin, Austin, TX 78712, USA}

\author{Naoki Yoshida}
\affiliation{Department of Physics, School of Science, University of Tokyo, Bunkyo, Tokyo 113-0033, Japan}
\affiliation{Kavli Institute for the Physics and Mathematics of the Universe (WPI), UT Institutes for Advanced Study, University of Tokyo, Kashiwa, Chiba 277-8583, Japan}

\author{Yuya Sakurai}
\affiliation{Department of Physics, School of Science, University of Tokyo, Bunkyo, Tokyo 113-0033, Japan}

\author{Michiko S. Fujii}
\affiliation{Department of Astronomy, School of Science, University of Tokyo, Bunkyo, Tokyo 113-0033, Japan}

\begin{abstract}
We perform a set of cosmological simulations of early structure formation incorporating baryonic streaming motions. 
We present a case where a significantly elongated gas cloud with $\sim\!10^4$\,solar mass ($\msun$) is formed in a pre-galactic ($\sim\!10^7\,\msun$) dark halo. 
The gas streaming into the halo compresses and heats the massive filamentary cloud to a temperature of $\sim\!10,000$\,Kelvin. 
The gas cloud cools rapidly by atomic hydrogen cooling, and then by molecular hydrogen cooling down to $\sim\!400$\,Kelvin. 
The rapid decrease of the temperature and hence of the Jeans mass triggers fragmentation of the filament to yield multiple gas clumps with a few hundred solar masses.
We estimate the mass of the primordial star formed in each fragment by adopting an analytic model based on a large set of radiation hydrodynamics simulations of protostellar evolution. 
The resulting stellar masses are in the range of $\sim\!50$--$120\,\msun$.
The massive stars gravitationally attract each other and form a compact star cluster. 
We follow the dynamics of the star cluster using a hybrid $N$-body simulation. We show that massive star binaries are formed in a few million years through multi-body interactions at the cluster center.
The eventual formation of the remnant black holes will leave a massive black hole binary, which can be a progenitor of strong gravitational wave sources similar to those recently detected by the Advanced Laser Interferometer Gravitational-Wave Observatory (LIGO).
\end{abstract}


\section{Introduction}
\label{sec:intro}

Recent detection of gravitational waves from binary black holes (BHs) by the Advanced Laser Interferometer Gravitational-Wave (GW) Observatory (LIGO) opened the new era of gravitational wave astronomy. 
The origin and evolution of such binary BHs with masses of several tens of solar masses are largely unknown, but it is thought that they are the remnants of stellar populations formed in low-metallicity environments. 
It has been also suggested that binary systems consisting of massive primordial stars can be progenitors of  gravitational wave sources.

There has been significant progress in the theoretical study of primordial star formation \citep[see][for recent reviews]{greif15,barkana16}.
Three-dimensional simulations with a fully cosmological set-up have been used to study in detail the formation process and physical properties of the first stars, galaxies, and black holes \citep[e.g.][]{hirano14,wise14,chon16}.
Generating realistic initial conditions is the key for such cosmological simulations.
\cite{tseliakhovich10} identify an important physical effect caused by baryonic supersonic motions relative to dark matter (DM) in the early Universe.
The initially supersonic motions quickly decay as the Universe expands, but cause considerable effects on early structure formation \citep[see][for a review]{fialkov14}.
The baryon fraction in small-mass dark halos is reduced \citep{naoz13}, gas condensation and subsequent star formation are suppressed or delayed \citep{greif11sv,stacy11sv}, and even the abundance of DM halos is affected \citep{dalal10,tseliakhovich11,fialkov12,naoz12,bovy13}.
Several important observational signatures are expected to be imprinted in large-scale structure, such as the large angular-scale fluctuation in the 21 cm intensity distribution \citep{dalal10,mcquinn12,visbal12,barkana13,fialkov13,tanaka16}, B-mode polarization of the cosmic microwave background \citep[CMB;][]{ferraro12}, and the number density of the low-mass satellite galaxies in the Milky Way \citep{bovy13}.

A number of authors have performed cosmological simulations in order to study the impact of the baryonic streaming motions. 
\cite{maio11} and \cite{naoz12} studied the formation of small mass dark halos and their gaseous contents, and \cite{o'leary12} and \cite{richardson13} studied the formation of star-forming gas clouds. 
\cite{greif11sv} showed that primordial star formation is delayed significantly and that the so-called minimum halo mass for gas collapse is raised by a factor of a few compared to the case without the streaming motions.
The global delay of star-formation due to the streaming motions causes both positive and negative feedbacks on the subsequent star-formation through fluctuations in the ultraviolet \citep[e.g.][]{haiman96uv} and X-ray background radiation \citep[e.g.][]{machacek03}.
The relative motions can cause the spatial offset between the baryon and DM density fluctuations. 
An intriguing possibility is that gas-deficient DM halos and DM-free gas clumps are formed, of which the latter can evolve into globular clusters, whereas the former can be progenitors of ultra-faint dwarf galaxies \citep{naoz14,popa16}.
Clearly, the early baryonic streams can generate a variety of objects from stars to star clusters and (dark) galaxies. 
It is thus important to follow the formation of individual early objects in detail using three-dimensional simulations.

We use cosmological hydrodynamics simulations with primordial chemistry and radiation transfer to study in detail the effect of baryonic streaming motions on the formation of the first stars.
Earlier in \cite{hirano17svI}, we presented several cases where supermassive stars are formed under very large baryonic streaming motions.
Primordial supermassive stars with masses $10^4$--$10^5\,\msun$ are formed in a manner similar to the turbulent core collapse model of present-day massive star formation \citep[e.g.][]{mckeetan02}.
In the present paper, we investigate other cases with low-to-moderate streaming velocities.
We find an interesting case where a large filamentary gas cloud is formed, which then fragments to yield multiple gas clouds.
Such filament fragmentation has not been seen in previous simulations without baryonic streaming motions, but could actually be one of the characteristic cases with realistic cosmological initial conditions because the streaming velocities assumed in the present study are not very large and correspond only to 2$\sigma$ fluctuations.
Finally, we follow the dynamical evolution of a cluster of stars formed via the filament fragmentation. 
We show that close-binary systems of massive stars are formed through multi-body interactions between the cluster member stars.

The rest of the paper is organized as follows.
We begin by describing the calculation methods in Section~\ref{sec:method}.
Section~\ref{sec:res} shows the results of cosmological simulations with different initial streaming velocities.
Section~\ref{sec:dis} discusses and summarizes the dependence of first star formation on the intrinsically generated streaming velocities.

\section{Numerical methodology}
\label{sec:method}

\begin{deluxetable*}{lcclrlccllc}[t!]
\tablecaption{Summary of cosmological simulations}
\tablehead{
\colhead{Identifier} &
\colhead{$v_{\rm bc}^{\rm rec}$} &
\colhead{$\sigma_8$} &
\colhead{$z$} &
\colhead{$R_{\rm V}$} &
\colhead{$M_{\rm tot,V}$} &
\colhead{$T_{\rm gas,V}$} &
\colhead{$f_{\rm gas,V}$} &
\colhead{$R_{\rm J}$} &
\colhead{$M_{\rm J}$} &
\colhead{$a/b-1$} \\
\colhead{} &
\colhead{($\sigma_{\rm bc}^{\rm rec}$)} &
\colhead{} &
\colhead{} &
\colhead{(pc)} &
\colhead{($10^7\,\msun$)} &
\colhead{(K)} &
\colhead{} &
\colhead{(pc)} &
\colhead{($10^3\,\msun$)} &
\colhead{}
}
\startdata
Run-No   & 0 & 0.8 & 34.6 &  26 & \hspace{2mm}0.016 & 1010 & 0.122 & \hspace{1.6mm}0.25 & \hspace{4.8mm}0.36 &  \hspace{1.6mm}1\\
\hspace{6.5mm}Low  & 1 & 0.8 & 27.9 &  76 & \hspace{2mm}0.22  & 2880 & 0.119 & \hspace{1.6mm}1.25 & \hspace{4.8mm}3.0 &  \hspace{1.6mm}5\\ 
\hspace{6.5mm}Med  & 2 & 0.8 & 20.1 & 293 & \hspace{2mm}3.35  & 9670 & 0.194 & 20 & \hspace{1.6mm}220 & 10\\ 
 & ... & ... & 24   & 125 & \hspace{2mm}0.51  & 4920 & 0.102 & ... & ... & ... \\ 
\hspace{6.5mm}High & 3 & 0.8 & 17.3 & 464 & \hspace{2mm}6.28  & 7430 & 0.204 & 79 & 2200 &  \hspace{1.6mm}4\\ 
 & ... & ... & 22   & 167 & \hspace{2mm}0.78  & 5750 & 0.084 & ... & ... & ... \\ 
\hspace{6.5mm}B    & 3 & 1.2 & 30.5 & 171 & \hspace{2mm}2.36  & 5320 & 0.085 & \hspace{1.6mm}2.5 & \hspace{3.2mm}28 &  \hspace{1.6mm}5\\ 
\hspace{6.5mm}A    & 3 & 2.0 & 49.4 &  93 & \hspace{2mm}2.76  & 6220 & 0.086 & \hspace{1.6mm}5.0 & \hspace{1.6mm}160 &  \hspace{1.6mm}0\\ 
\enddata
\tablecomments{
Column 1: identification of simulation;
Column 2: relative streaming velocity normalized by the root-mean-square value $\sigma_{\rm bc}$; 
Column 3: a root-mean-square density fluctuation in a sphere of radius $8\,h^{-1}$\,Mpc; 
Column 4: redshift when the gas number density at the collapsing center reaches $\nhcen = 10^6\,\cc$; 
Columns 5 to 8: radius, total mass, gas temperature, and gas fraction at the virial scale;
Columns 9 and 10: radius and mass at the Jeans scale; and
Column 11: ellipticity of collapsing gas cloud $a/b - 1$, where $a$ and $b$ are the major and minor axes.
The second rows for Run-Med and High shows results for when the gas collapse can occur if the collapse suppression via halo mergers was to be ignored.
The bottom two are results in \cite{hirano17svI}, which leave a supermassive protostar with $\sim\!10^5\,\msun$.
}
\label{tab1}
\end{deluxetable*}

We perform a set of cosmological simulations that incorporate the early baryonic streaming motions.
We use a hierarchical zoom-in technique to achieve sufficiently high spatial resolution to follow  the hierarchical assembly of small-mass dark matter halos and the formation of the star-forming gas cloud within them.
The parent cosmological simulation has a volume of $L_{\rm box} = 10\,h^{-1}$\,comoving megaparsec (cMpc) on a side, in which we select zoom regions with $L_{\rm zoom} = 0.3\,h^{-1}\,{\rm cMpc}$.
In the high-resolution regions, the particle masses of dark matter and gas components are $m_{\rm DM} = 16.4$ and $m_{\rm baryon} = 3.0\,\msun$, respectively.
We use the publicly available code {\sc music} \citep{hahn11} to generate the cosmological initial conditions at redshift $z_{\rm ini} = 499$.
Note that the initial epoch is chosen so that the high-order perturbations are accurately reproduced.\footnote{\citet{o'leary12} suggested setting high initialization redshifts with $z_{\rm ini} > 200$.}
We adopt the standard $\Lambda$-cold dark matter ($\Lambda$-CDM) cosmology with total matter density $\Omega_{\rm m} = 0.3086$, baryon density $\Omega_{\rm b} = 0.04825$, dark energy density $\Omega_{\Lambda} = 0.6914$ in units of the critical density, Hubble constant $h = 0.6777$,  density fluctuation amplitude $\sigma_8 = 0.8288$, and  primordial index $n_{\rm s} = 0.9611$ \citep{PLANCK13XVI}. 
The initial ionization fraction is $x_{\rm e} = 6.88 \times 10^{-4}$ \citep{seager99,seager00,wong08}.

The streaming motions are realized in a straightforward manner by simply adding a constant uniform velocity along one axis to the initial conditions of our zoomed simulations.
This procedure is valid because the distribution of the streaming velocity is coherent over a length of a few mega-parsecs, which is larger than a typical region that contains target first galaxy halos of our interest \citep{tseliakhovich10}.
Furthermore, the initial streaming direction can be set arbitrarily because the cosmological supersonic motion is not correlated with the local overdensity \citep{ahn16}.
In practice, we introduce the initial relative velocity $v_{\rm bc}$ (``bc'' representing ``baryon-cold dark matter'') into the otherwise standard initial conditions.
We do not consider spatial offset between the two components at $z_{\rm ini}$ because it is negligibly small at the initial redshift \citep{naoz12}.
We generate four cosmological initial conditions with the same phase for the density field but with different $v_{\rm bc}$ normalized by the root-mean-square value, $\sigma_{\rm bc}^{\rm rec} = 30\,\kms$, at the epoch of cosmological recombination $z_{\rm rec} = 1089$.
We adopt the following values and dub the four runs as $v_{\rm bc}^{\rm rec} / \sigma_{\rm bc}^{\rm rec} = 0$ (Run-No), 1 (Low), 2 (Med), and 3 (High).
Table~\ref{tab1} summarizes the parameter sets of the cosmological initial conditions.
We also refer to the simulations in \cite{hirano17svI}, for which the initial density fields are generated with $\sigma_8 = 1.2$ (Run-B) and $2.0$ (A).

The cosmological simulations are performed using the parallel $N$-body/smoothed particle hydrodynamics (SPH) code {\sc gadget-2} \citep{springel05gadget2} suitably modified for primordial star formation \citep{yoshida07,yoshida08,hirano15}.
We employ a hierarchical refinement technique to follow the gas cloud collapse as in \cite{hirano17svI}, with refinement criterion that the local Jeans length is always resolved by $15$ times the local smoothing length by progressively increasing the spatial resolution using the particle-splitting technique of \citet{kitsionas02}.
The minimum mass of gas particles becomes $m_{\rm baryon,min} = 1.6 \times 10^{-4}\,\msun$, which is enough to avoid the artificial delay of gas collapse \citep{greif11sv}.
We stop the simulations when the maximum hydrogen number density reaches $\nh = \rho / m_{\rm H} = 10^{12}\,\cc$ where $m_{\rm H}$ is the proton mass.

\section{Hydrodynamical Simulations}
\label{sec:res}

\begin{figure*}[t!]
\begin{center}
\includegraphics[width=1\textwidth]{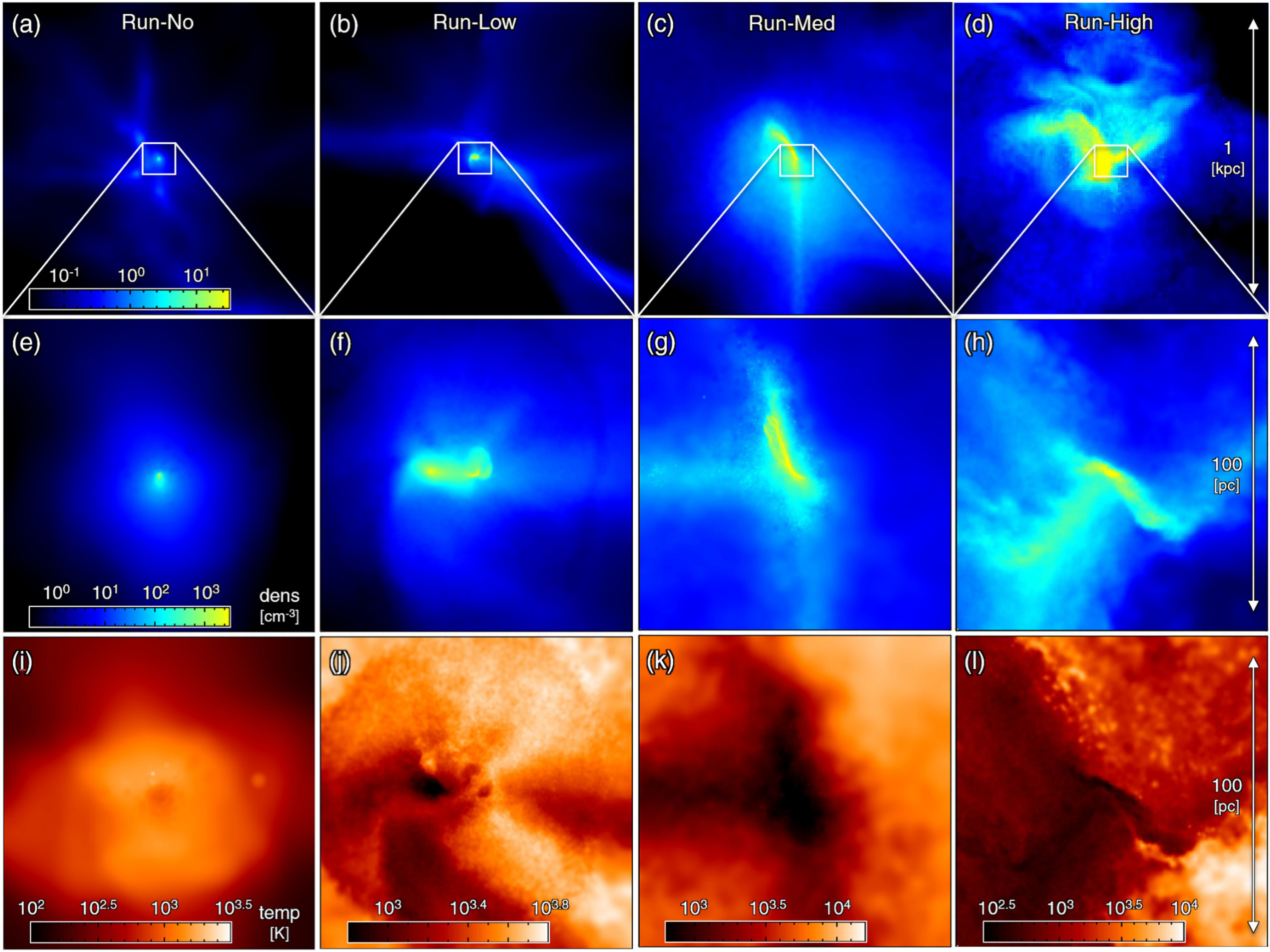}
\caption{
Projected structure of the primordial star-forming cloud in Run-No (at $z = 34.6$), Low ($27.9$), Med ($20.1$), and High ($17.3$): gas number density in a region of $1$\,kpc on a side (panels (a) to (d)), and gas number density ((e) to (h)) and gas temperature ((i) to (l)) in a region of $100$\,pc on a side.
The direction of the initial supersonic stream between the gas and dark matter components is aligned to the horizontal axis (from left to right) in the figure.
}
\label{fig1}
\end{center}
\end{figure*}

We confirm that the previously known effects of the streaming motions are reproduced in our simulations. 
Namely, the gas cloud collapse is delayed and the host halo mass increases until the gas cloud finally collapses.
Table~\ref{tab1} summarizes the properties of the star-forming gas clouds in our four runs at the respective time when gravitational collapse occurs.
In addition to these expected results, we also find significant deformation of the collapsing gas clouds.
Figure~\ref{fig1} shows clearly that the cloud appears nearly spherical but is elongated to be filamentary with increasing streaming velocity (from left to right panels).

\subsection{Formation of Star-forming Gas Cloud}
\label{sec:res_delay}

\begin{figure}[t!]
\begin{center}
\includegraphics[width=0.45\textwidth]{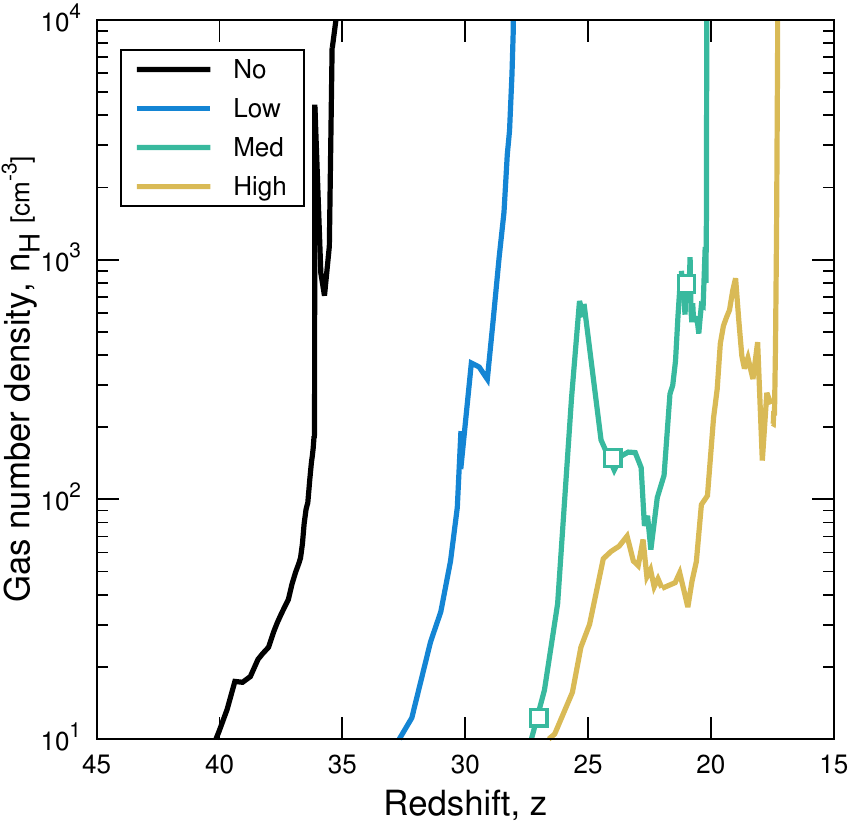}
\caption{
Time evolution of the maximum gas number density as a function of redshift in Run-No (black), Low (blue), Med (green), and High (yellow), respectively.
The squares on the line for Run-Med indicate the three output times, $z = 27$, $24$, and $21$, used in Figure~\ref{fig3} to show the structural evolution.
}
\label{fig2}
\end{center}
\end{figure}

\begin{figure}[t!]
\begin{center}
\includegraphics[width=0.45\textwidth]{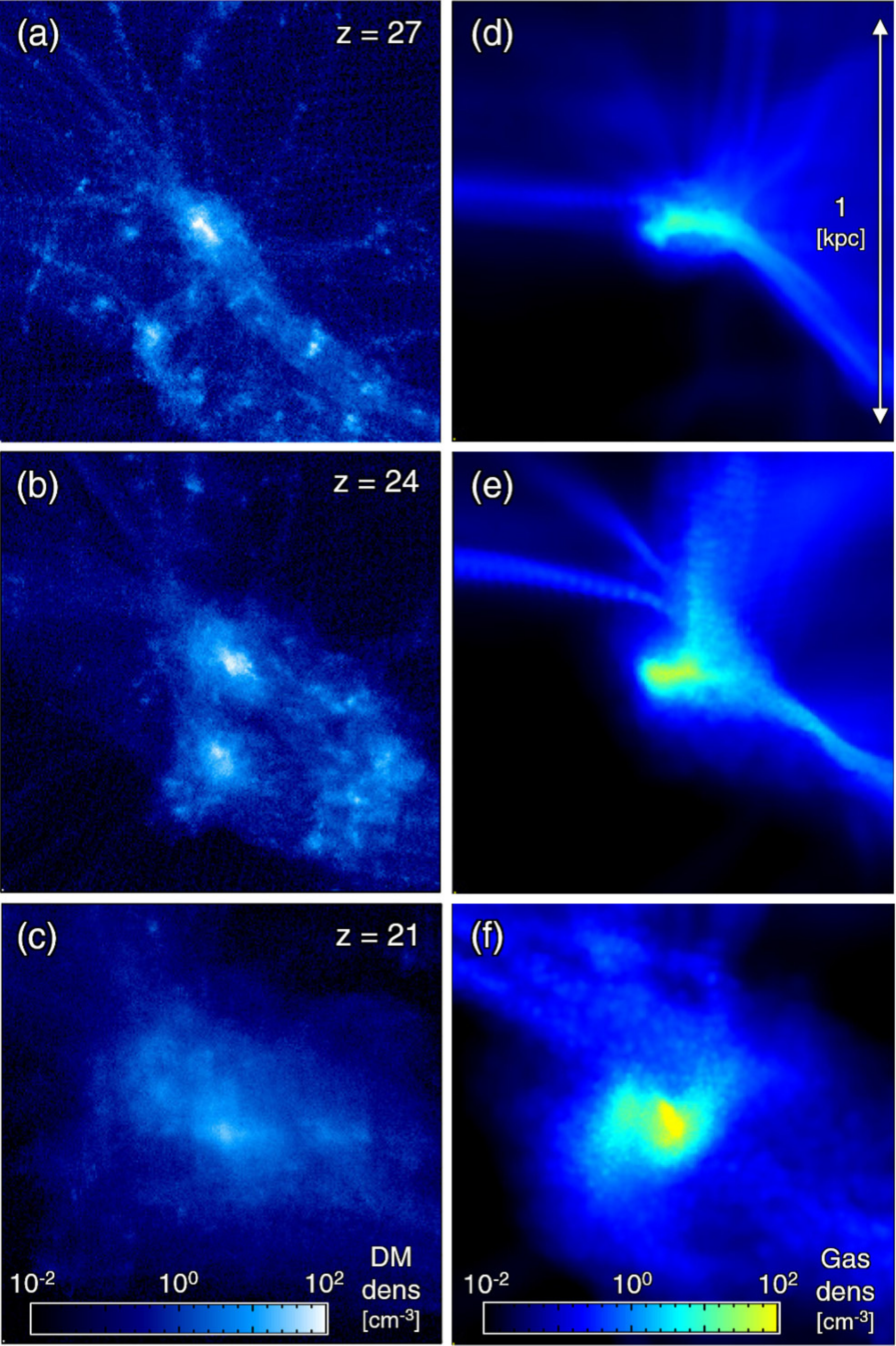}
\caption{
Cross-sectional density distributions in a region of $1$\,kpc on a side around the primordial star-forming cloud in Run-Med: dark matter (panels (a) to (c)) and gas ((d) to (f)) components, respectively.
The three rows show the structure at three epochs indicated by the squares in Figure~\ref{fig2}: $z = 27$ (panels (a) and (d)), $24$ ((b) and (e)), and $21$ ((c) and (f)), respectively.
}
\label{fig3}
\end{center}
\end{figure}

A major effect of the streaming velocity on structure formation is suppression or delay of the gas condensation in small-mass dark matter halos. 
Figure~\ref{fig2} shows the time variation of the maximum gas density in each run. 
The collapse epoch is systematically delayed to lower redshifts with increasing streaming velocity.
As an example, let us compare Figures~\ref{fig1}(b) and \ref{fig3}(d).
The gas cloud has already collapsed ($\nh \ge 10^6\,\cc$) in the former (Run-Low), but the latter (Run-Med) reaches just only $\nh \sim 10\,\cc$ at similar epochs $z = 27$--$28$ (see Figure~\ref{fig2}).
Note that the dark matter distributions are nearly the same at any given epoch in Run-No to Run-High, regardless of the baryonic streaming velocities.  
Figure~\ref{fig1}(c) shows that the halo gas is compressed and vertically elongated at $z = 17.3$. 
Then the host halo mass is already greater than $10^7\,\msun$, and its deep gravitational potential can trap the gas streams inflowing from the left in the figure.
In Run-High, the gas distribution is highly disturbed within the virial radius of $R_{\rm V} = 464$\,pc (Figure~\ref{fig1}d).
The subsequent gravitational collapse of the central gas cloud proceeds in a dynamical and complicated manner (Figure~\ref{fig1}h), as has been also reported in \citet{hirano17svI}.

The overall impact of the streaming motions weakens with time because the relative velocity decays with decreasing redshift as
\begin{equation}
v_{\rm bc}(z) = v_{\rm bc}^{\rm rec} \left( \frac{1+z}{1+z_{\rm rec}} \right) \, .
\label{eq:v_bc}
\end{equation}
However, in addition to the overall delay of collapse, halo mergers continue disturbing the gas collapse  
after $z = 25$ in Run-Med and High (Figure~\ref{fig2}).
A number of minihalos fall in along dark matter ``filaments'' and are accreted onto the central halo (Figures~\ref{fig3}(a) to (c)).
Through the dynamical relaxation process, halo mergers prevent the gas from cooling and condensing, until 
the gas temperature reaches $\sim 10,000$ K and efficient atomic hydrogen cooling operates.
To summarize, the fast gas streams delay gas condensation, and hence star formation in small dark halos, and the gas in growing dark halos is kept dynamically ``hot'' for a few tens of millions of years. 
Star-forming gas clouds are formed only after the virial halo temperature exceeds the so-called
atomic cooling threshold.

\begin{figure}[t!]
\begin{center}
\includegraphics[width=0.45\textwidth]{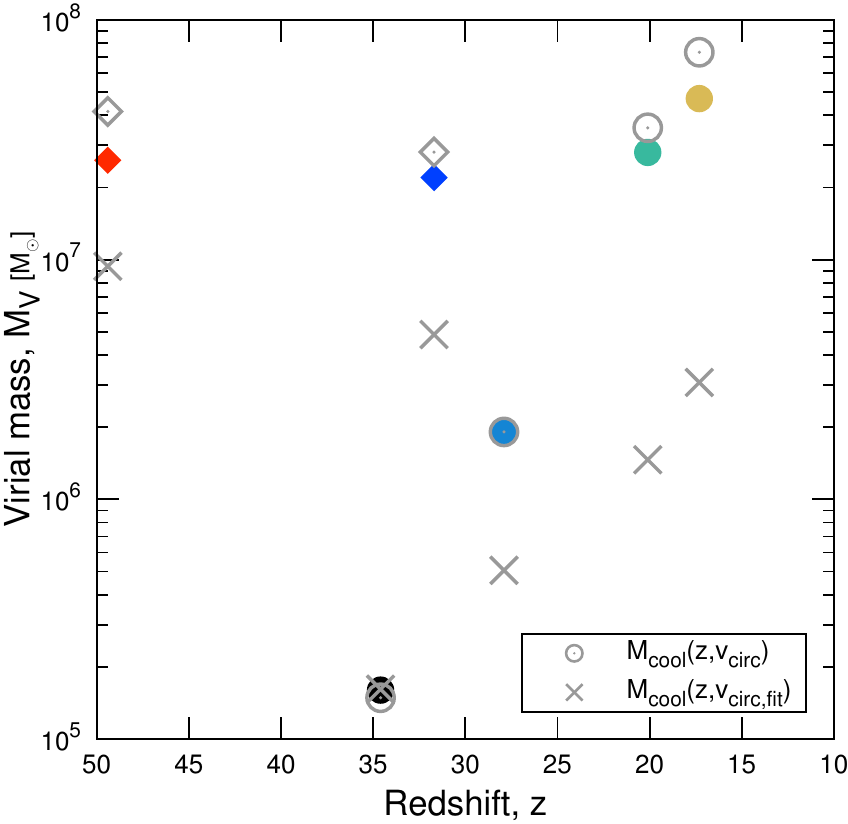}
\caption{
We plot the virial halo mass against collapse epoch (redshift) when the central gas density reaches $\nhcen = 10^6\,\cc$.
The open circles show the corresponding threshold masses obtained from Equation~(\ref{eq:M_cool}).
The crosses indicate the results from the fitting function (equation~\ref{eq:v_circ,fit}) with the parameters $v_{\rm circ,0}=3.7\,\kms$ and $\alpha = 4.0$. 
The open and filled diamonds, plotted for comparison, are the results for Run-A and Run-B in \cite{hirano17svI}.
}
\label{fig4}
\end{center}
\end{figure}

\begin{figure}[t!]
\begin{center}
\includegraphics[width=0.45\textwidth]{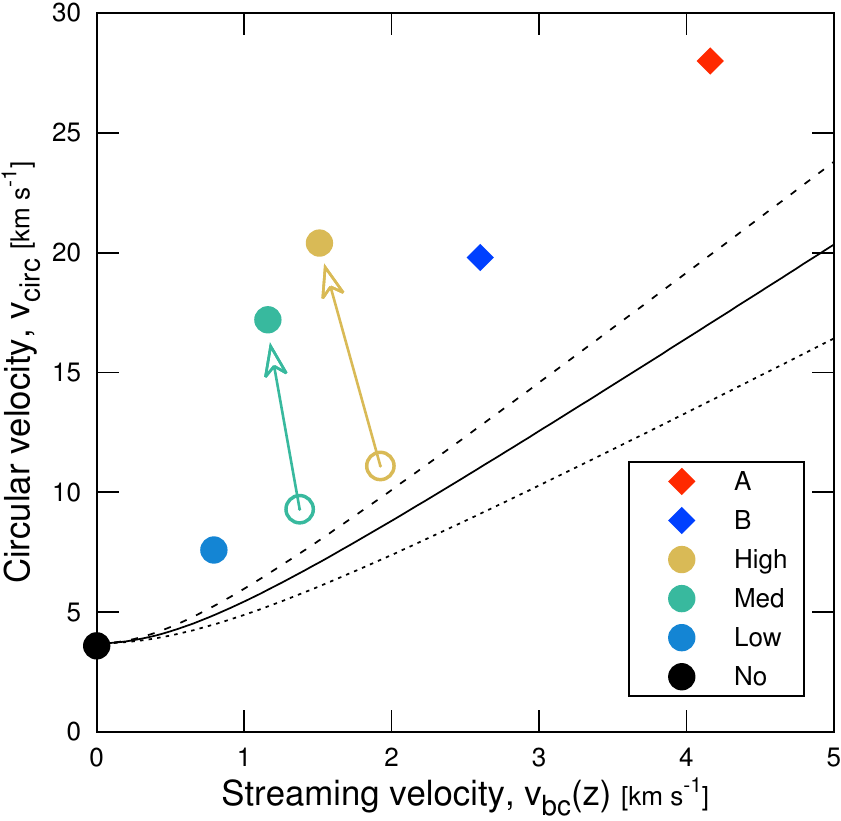}
\caption{
We plot the circular velocities of the host halos when the central gas density reaches $\nhcen = 10^6\,\cc$ as a function of the bulk streaming velocity $v_{\rm bc}(z)$.
The open circles indicate the results at $z = 24$ in Run-Med and $z = 22$ in Run-High when the cloud collapse is disturbed by minihalo mergers (see Figure~\ref{fig2}).
The dashed, solid, and dotted lines represent fitting functions (equation~\ref{eq:v_circ,fit}) with the parameters $v_{\rm circ,0}=3.7\,\kms$ and $\alpha = 4.7$, $4.0$, and $3.2$, respectively.
}
\label{fig5}
\end{center}
\end{figure}

With the dynamical effect taken into account, the minimum halo mass for gas cloud formation is significantly
larger than in the case without streaming motions.
Figure~\ref{fig4} shows that the virial halo masses at the onset of gas collapse (filled symbols) agree well with the cooling threshold (open symbols) derived from the virial halo temperature given by
\begin{eqnarray}
  M_{\rm cool} \approx 4.5 \times 10^5\,\msun\,\left( \frac{v_{\rm circ}}{4\,\kms} \right)^3 \left( \frac{1+z}{21} \right)^{-3/2} \, ,
  \label{eq:M_cool}
\end{eqnarray}
where $v_{\rm circ} = \sqrt{GM_{\rm V}/R_{\rm V}}$ is the circular velocity, $G$ is the gravitational constant, and $M_{\rm V}$ and $R_{\rm V}$ are the virial mass and radius \citep[e.g.][]{barkana01}.
Interestingly, a naive model that accounts for the effect of the streaming motions does not reproduce our results.
\citet{fialkov12} proposed that the increase of the threshold mass can be formulated by using the ``effective'' velocity 
\begin{eqnarray}
  v_{\rm circ,fit}(z) = \sqrt{ v_{\rm circ,0}^2 + [\alpha v_{\rm bc}(z)]^2 }.
  \label{eq:v_circ,fit}
\end{eqnarray}
With appropriate choice of the two parameters, $v_{\rm circ,0}$ and $\alpha$, the simulation results of \cite{stacy11sv} and \cite{greif11sv} are reproduced \citep{fialkov12}.
However, this model does not describe well the results of our simulations (compare crosses and filled symbols in Figure~\ref{fig4}).
This can be easily seen in Figure~\ref{fig5}, where we compare the circular velocities of halos at the onset of gas collapse as a function of $v_{\rm bc}(z)$ (equation~\ref{eq:v_bc}).
Data for Run-No and Low can be fitted by $v_{\rm circ,fit}(z)$ with parameters $v_{\rm circ,0}=3.7\,\kms$ and $\alpha = 3.2$--$4.7$.
Contrastingly, in Run-Med and Run-High, the host halos have grown significantly through mergers to over the estimated cooling threshold mass. 
We note that the halo circular velocity at $z = 24$ in Run-Med (indicated by the open circle) is actually close to the model prediction. 
While the gas cloud is contracting at $z = 24$, a series of halo mergers continuously disturb, and the final collapse occurs late at $z = 20$ (see also Figure~\ref{fig2}). 
This is also found in Run-High, as shown in Figure~\ref{fig5}, and essentially the same features are found in recent simulations by \citet{schauer17}. 
With large streaming velocities, the critical condition for gas cooling and collapse cannot be determined by the host halo mass (or circular velocity). 
Direct numerical simulations that follow the assembly of dark matter halos and the gas thermal evolution are necessary.

The gas fraction in the host halos, $f_{\rm gas,V}$, decreases due to the gas escape by the streaming motion but increases the prolonged delay of collapse (Table~\ref{tab1}).
We find that the gas fraction actually decreases with increase in the initial relative velocity, $f_{\rm gas,V} = 0.122$, $0.119$, $0.102$, and $0.084$ for Run-No, Low, Med ($z = 24$), and High ($z = 22$), respectively.
These fractions are lower than the cosmic average, $\Omega_{\rm b} / \Omega_{\rm m} = 0.156$, because of the escape of gas density fluctuation from the DM structure.
On the contrary, in fact, the values in Run-Med and High, $f_{\rm gas,V} = 0.194$ and $0.204$, overcome the cosmic average and slightly increase with increase in the initial relative velocity.
In these cases, the gas component within the virial radius is gravitationally trapped and the baryonic bulk motion works as external pressure to compress the gas component to the host halo.
As a result, the gas clouds tend to be ``squashed'' vertically with respect to the streaming motions (see Figures~\ref{fig1}g and \ref{fig1}h).
This can be seen typically at low redshifts ($z \sim 15$--$25$) when the host halos can gravitationally trap the streaming gas efficiently.
In Run-A and B of \cite{hirano17svI} with the same initial streaming velocity but in high-density regions, the gas fractions are much lower than the cosmic mean, $f_{\rm gas,V} \approx 0.085$.

\subsection{Fragmentation of Elongated Cloud}
\label{sec:res_elon}

We stop the simulations when the gas density at the cloud center reaches $\nh = 10^{12}\,\cc$.
By the respective final epoch, the host halo has grown to about $10$--$100$ times that in Run-No (Table~\ref{tab1}).
In such a massive halo, the star-forming cloud is surrounded by a dense gas envelope. 
We find that the density distribution in Run-No is approximated by a power-law function as $\rho \propto R^{-2.2}$ \citep{omukai98}, but the other cases show strongly enhanced density structure at large radii.
The density profile steepens at the Jeans scale, $R_{\rm J}$ (Table~\ref{tab1}), inside which the 
gas cloud undergoes runaway collapse.
At the innermost region, the density profile is similar to the power-law shape of the cloud in Run-No.

\begin{figure}[t!]
\begin{center}
\includegraphics[width=0.45\textwidth]{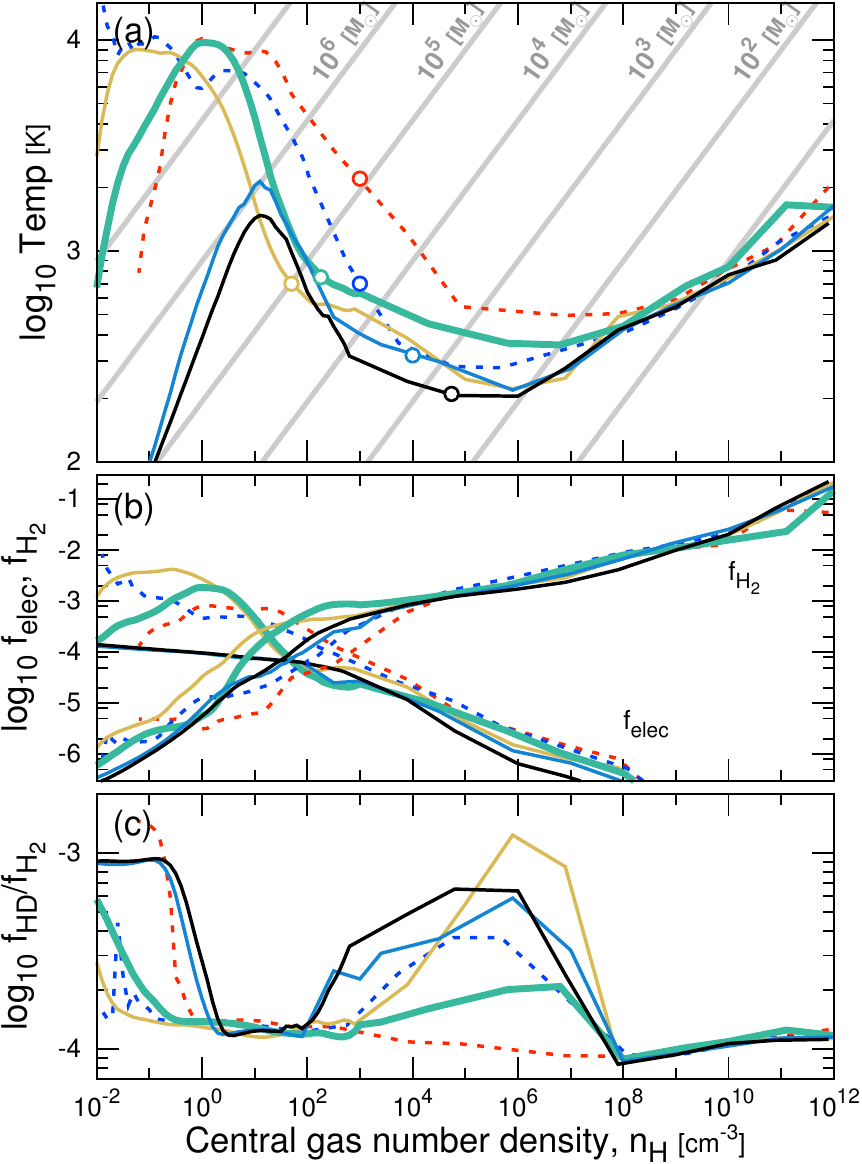}
\caption{
Thermal and chemical evolution of the collapsing gas clouds as a function of the gas number density: gas temperature (panel (a)), electron and H$_2$ fractions (b), and ratio of HD and H$_2$ fractions (c).
The solid lines show the results for Run-No (black), Low (blue), Med (green), and High (yellow), respectively.
The dashed lines show the results for Run-A (red) and B (blue) in \cite{hirano17svI}.
The circles mark the epochs when the gas clouds become Jeans unstable.
HD line emission becomes efficient in a primordial cloud, which has cooled by H$_2$ cooling from a highly ionized state with $\sim\!10^4$\,K \citep{yoshida07}.
Similar conditions are realized in Run-Med and High (panel (b)).
The necessary HD fraction to cool the gas is estimated to be $f_{\rm HD} / f_{\rm H_2} \sim 10^{-3}$ at $\nh < 10^6\,\cc$.
The HD fraction increases in all of our runs, but the gas temperature does not get sufficiently low due to enhanced compressional heating.
Thus HD cooling does not become efficient in Run-Med (panel (c)), whereas in Run-High, the central core temporarily cools to the CMB temperature floor, $T_{\rm CMB}(z) = 2.73(1+z) \approx 50$\,K at $z = 17.3$, but quickly gets to around $250$\,K owing to compression heating.
}
\label{fig6}
\end{center}
\end{figure}

Figure~\ref{fig6}(a) shows the thermal evolution of the collapsing gas clouds.
There are two major radiative cooling processes in the pristine gas.
Atomic hydrogen (H) cooling becomes efficient at $T \ge 8000$\,K \citep{barkana01}, and molecular hydrogen (H$_2$) line emission can further cool the gas down to $T \sim 200$\,K.
The gas cloud in Run-No cools by H$_2$-cooling when a sufficient amount 
of hydrogen molecules are formed at $\nh \sim 10\,\cc$.
In Run-Med and High with large streaming velocities, the gas temperature first reaches $\sim\!8000$\,K, above which atomic hydrogen cooling becomes efficient, and then the gas cools rapidly.

While condensing in the dark halo, the gas cloud becomes gravitationally unstable when its mass exceeds the Jeans mass, which is given by
\begin{equation}
M_{\rm J} \approx 10^3\,\msun\,\left( \frac{\cs}{1\,\kms} \right)^3 \left( \frac{\nh}{10^4\,\cc} \right)^{-1/2} \, ,
\label{eq:Mjeans}
\end{equation}
where $\cs$ is the sound speed of about a few\,$\kms$ in the collapsing primordial cloud.
Because the gas clouds are supported by thermal pressure and also by turbulent motions induced by the streaming motions, we evaluate the Jeans mass by replacing $\cs$ with $\{ \cs^2 + v_{\rm bc}^2(z) \}^{1/2}$.
Runaway gravitational collapse is triggered when the mass of the collapsing gas cloud overcomes this critical mass (circles in Figure~\ref{fig6}a).
After a brief period of gas condensation, the temperature increases again, but slowly, when compressional heating becomes effective (Figure~\ref{fig6}a).

The star-forming cloud transforms from spherical to having an elongated structure with increasing streaming
velocity (Figures~\ref{fig1}(e) to \ref{fig1}(h)).
We find that the filamentary clouds shrink radially at first, while the cloud ellipticity, $a/b-1$, where $a$ and $b$ is the major and minor axes, is roughly maintained (Table~\ref{tab1}). 
When the contracting filament, with its temperature decreasing owing to radiative cooling, becomes gravitationally unstable, it breaks up into multiple gas clumps.

Collapse and fragmentation of an elongated gas cloud can be well described by self-similar solutions of a cylindrical filament \citep[e.g.][]{inutsuka92,inutsuka97}.
A dense filament becomes unstable to axisymmetric perturbations of wavelength greater than about twice 
the filament diameter, when the line mass (gas mass per unit length) is close to the equilibrium value $\simeq\!2 c_{\rm s}^2/G$.
On the other hand, if the line mass exceeds the equilibrium value, the whole filament collapses along the major axis without fragmentation.
The former case is found in Run-Low, Run-Med, and Run-High in this study, whereas the latter case is realized in Run-B in \cite{hirano17svI}, where a large fraction of the initially filamentary cloud with $a/b-1 = 5$ is finally accreted onto a central protostar.

\begin{figure}[t!]
\begin{center}
\includegraphics[width=0.45\textwidth]{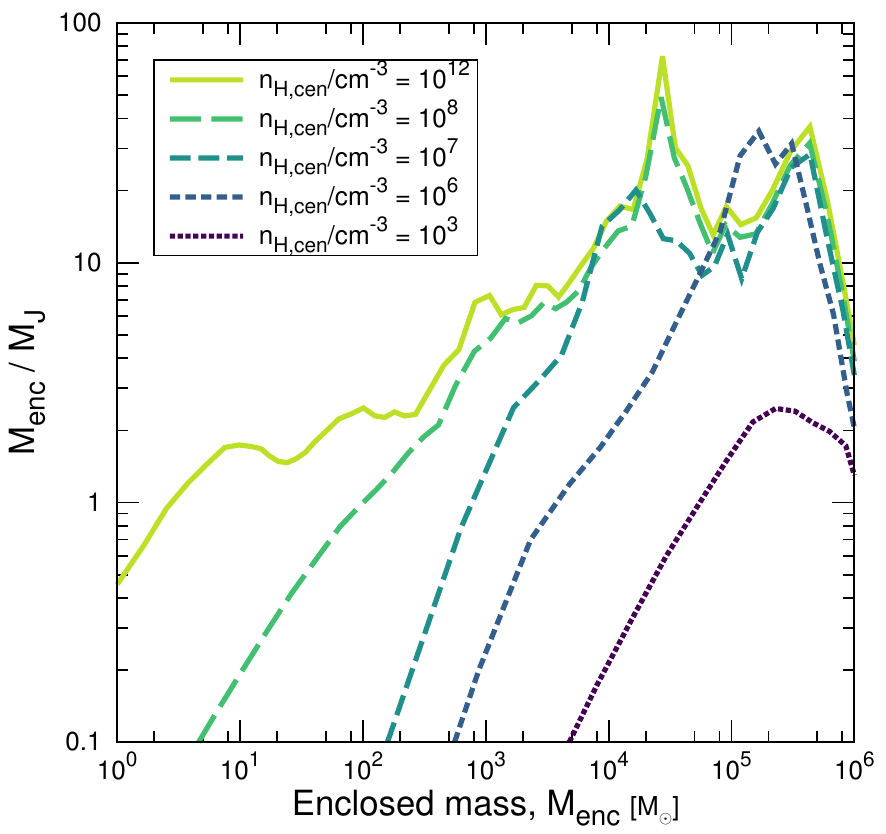}
\caption{
Gravitational instability of the collapsing gas cloud in Run-Med. 
We plot $M_{\rm enc} / M_{\rm J}$ as a function of the enclosed mass when the central gas number density reaches $\nhcen = 10^3$, $10^6$, $10^7$, $10^8$, and $10^{12}\,\cc$, respectively.
The baryonic streaming velocity in Run-Med is $v_{\rm bc}(z) = 2\sigma_{\rm bc}^{\rm rec}\,(1+z)/(1+z_{\rm rec}) = 1.16$\,$\kms$ at the collapse redshift $z = 20.1$.
}
\label{fig7}
\end{center}
\end{figure}

The cosmological streaming motions delay gas cloud formation but also affect the subsequent evolution of star-forming gas clouds in a complicated manner.
The cloud in Run-No collapses nearly spherically, and forms a single protostar at the center (Figure~\ref{fig1}e).
In Run-Low, an elongated structure appears when $\nhcen \sim 10^3\,\cc$ and two fragments with a separation about $10\,{\rm pc} \sim 2L_{\rm J}$ (Jeans length) are formed when gravitational instability sets in at $\nhcen \sim 10^4\,\cc$ (Figure~\ref{fig1}f).
In Run-Med and Run-High, large filamentary structures are formed and yield a number of fragments when $\nhcen \sim 10^5$--$10^6\,\cc$.
In Figure~\ref{fig7}, we show the time evolution of $M_{\rm enc} / M_{\rm J}$ in Run-Med.
After the first unstable cloud with $\sim\!10^5\,\msun$ forms when $\nhcen = 10^3\,\cc$, the inner filamentary structure with mass $\sim\!10^4\,\msun$ becomes highly gravitationally unstable.
The densest part the filamentary structure has a density of $\sim\!10^6\,\cc$, and its collapse occurs just after the so-called loitering point.

\subsection{Formation of Massive First Star Cluster}
\label{sec:res_cluster}

What is the final fate of the cluster of massive gas clumps found in Run-Med?
An important question is whether most of, if not all, of the fragments collapse to yield stars before the very first star grows and causes radiation feedback effects in its surroundings.
The typical evolution timescale from the protostar formation to the zero-age main sequence is $10^4$--$10^5$\,years \citep[see figure~1 in][]{hirano17acc}.
This should be compared with the free-fall time of a typical cloud (fragment),
\begin{eqnarray}
t_{\rm ff} = \sqrt{ \frac{3 \pi}{32 G \rho} } = 5.2 \times 10^4\,{\rm yr}\,\left( \frac{\nh}{10^6\,\cc} \right)^{-1/2} \, .
\label{eq:t_ff}
\end{eqnarray}
Note that the dense fragments can also shield themselves against external radiation. 
Hence, each gas cloud, or at least its densest part, likely continues contracting even under the influence of radiation from nearby stars forming at the same time.
We thus expect that most of the fragments with densities greater than $\sim\!10^6\,\cc$, as found in our simulations, can actually collapse and form stars before being disrupted significantly by the radiation feedback from the other nearby stars.

\begin{figure}[t!]
\begin{center}
\begin{tabular}{c}
\includegraphics[width=0.9\columnwidth]{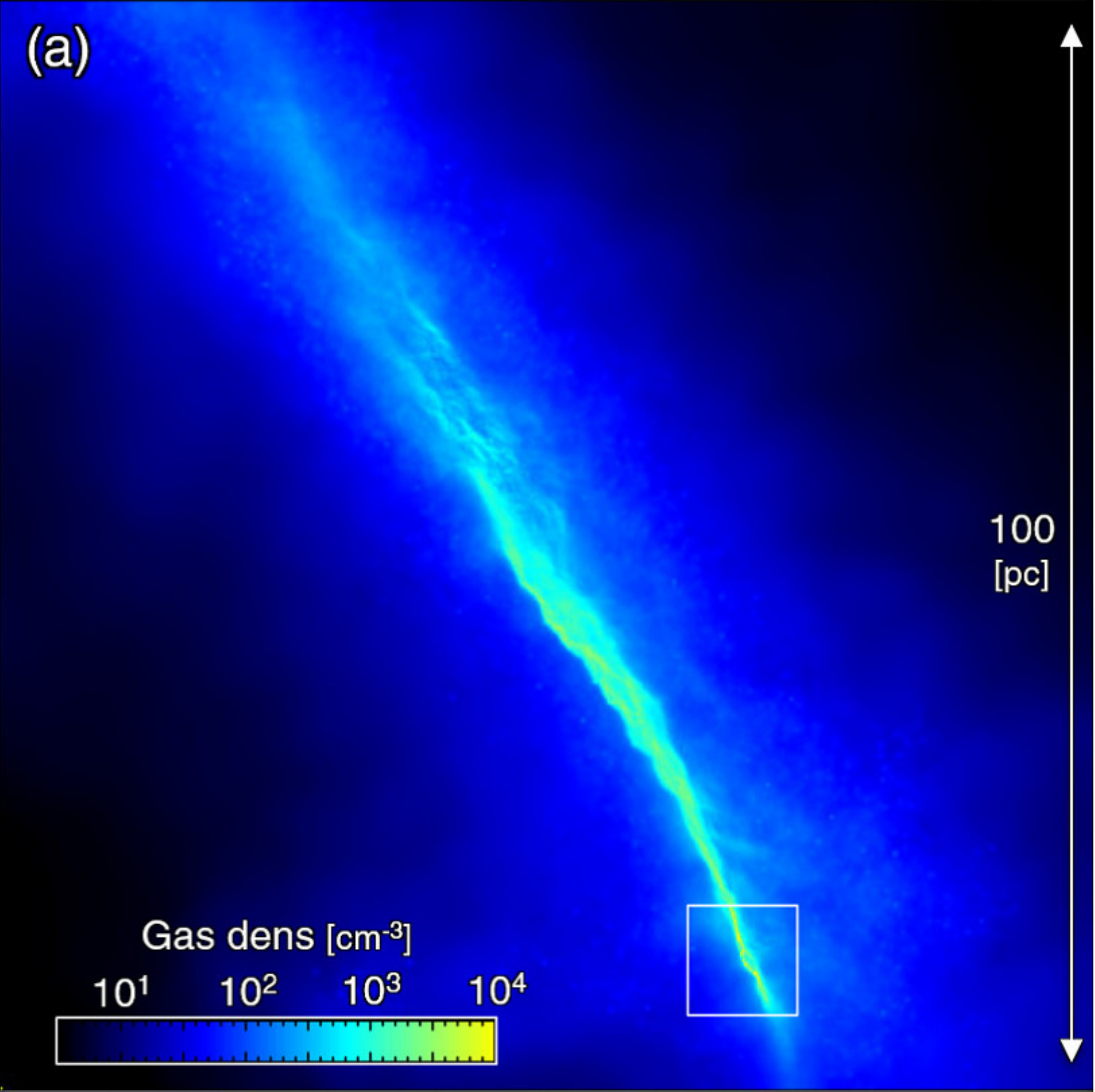}\\
\includegraphics[width=0.9\columnwidth]{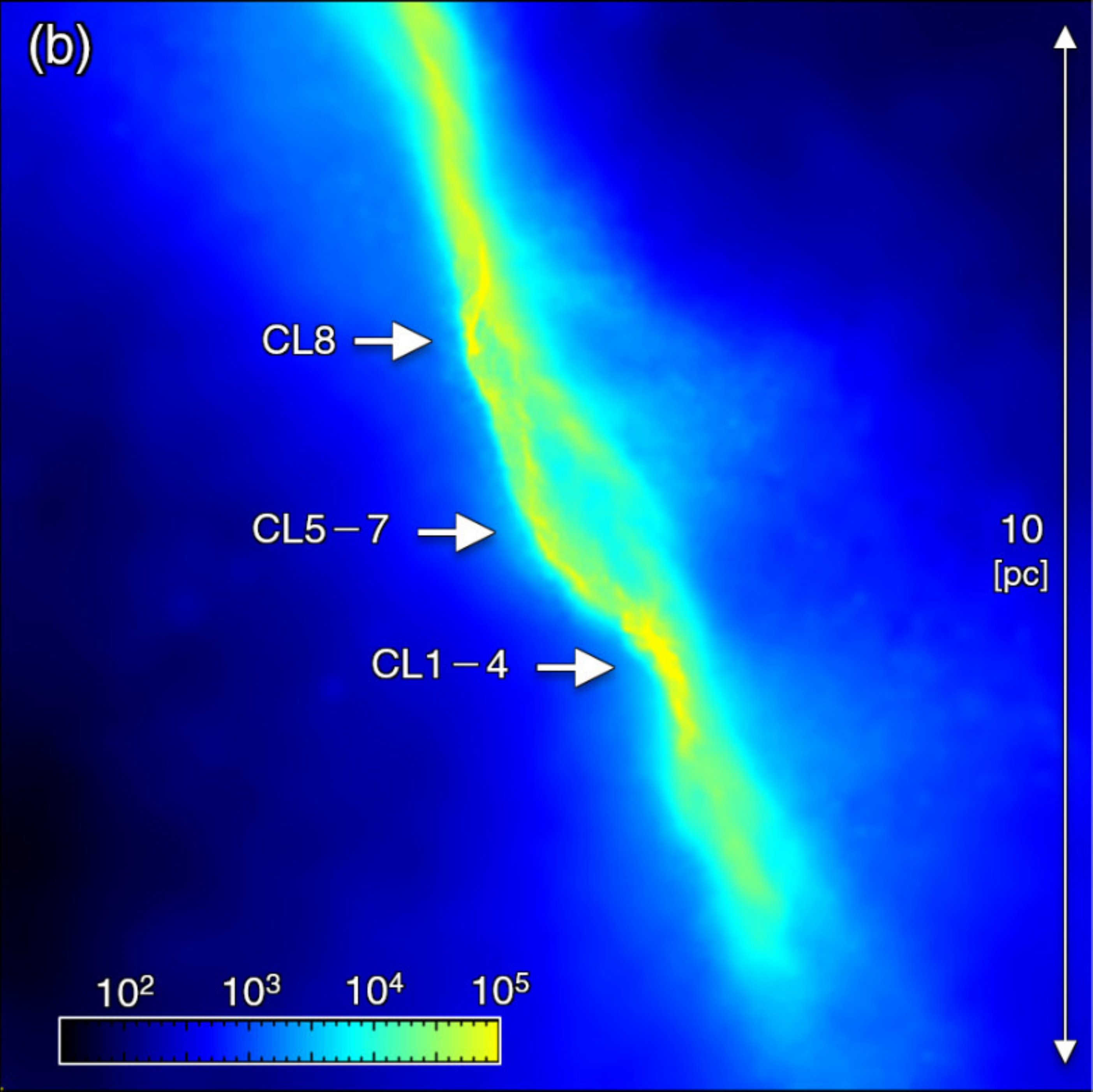}
\end{tabular}
\caption{
Cross-sectional gas density distribution around the filament including one of the early-forming clumps (CL1) in Run-Med.
We show a region of $100$\,pc (panel a) and $10$\,pc (b) on a side. 
We use a snapshot at $40$\,kyr after CL1 is formed.
}
\label{fig8}
\end{center}
\end{figure}

Figure~\ref{fig8} shows the fine structure of the filamentary cloud in Run-Med.
The bottom panel shows the most massive part, including a few fragments. 
The region within the density contour with $\nh = 10^6\,\cc$ has $5.8$\,pc in length and $0.2$\,pc in width.
Note the Jeans length $L_{\rm J} \simeq 0.4$\,pc for $T = 400$\,K and $\nh = 10^6\,\cc$.
To study further the evolution of the density fluctuations in the filament, we continue the simulation by introducing the technique of \cite{hirano17acc}.
We followed the evolution for $40,000$\,years after the first clump (CL1) formation, and found that a total of eight clumps were formed in the same filament.

\begin{deluxetable}{lrrrrrrrr}[t!]
\tablecaption{Properties of gravitationally unstable gas clumps}
\tablehead{
\colhead{} & \colhead{CL1} & \colhead{CL2} & \colhead{CL3} & \colhead{CL4} & \colhead{CL5} & \colhead{CL6} & \colhead{CL7} & \colhead{CL8}
}
\startdata
$M_{\rm frag}/\msun$ & 240 & 180 & 400 & 410 & 240 & 280 & 350 & 170 \\
$M_{\rm star}/\msun$ &  74 &  59 & 111 & 113 &  74 &  84 & 100 &  56 \\
\enddata
\tablecomments{
The gravitationally unstable gas mass around the clump, $M_{\rm frag}$, and the stellar mass estimated by a fitting function, $M_{\rm star}$ \citep[eq.~17 in][]{hirano14}.
}
\label{tab2}
\end{deluxetable}

\begin{figure}[t!]
\begin{center}
\includegraphics[width=0.45\textwidth]{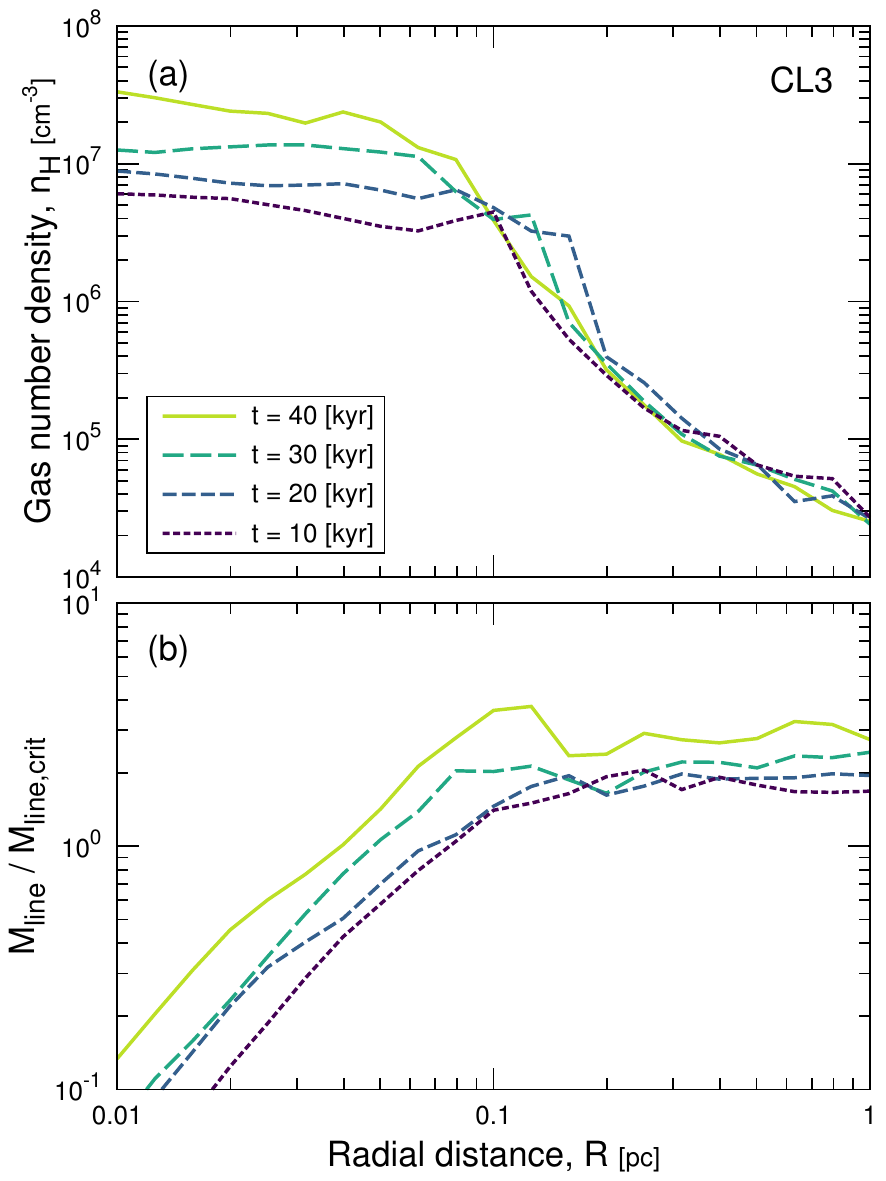}
\caption{
We plot the radially averaged density profile around the clump CL3 formed in the main filament of Run-Med: density distribution (panel (a)) and gravitational instability (b) at $t = 10$, $20$, $30$, and $40$\,kyr.
}
\label{fig9}
\end{center}
\end{figure}

To identify gravitationally unstable fragments in the filament, we perform a series of procedures as follows. 
We first identify the filament as a cloud of gas particles with densities greater than $10^5\,\cc$.
\begin{enumerate}
\item We define the direction of the filament by a vector pointing from one side of the filament to the other. 
\item We divide the filament into a series of bins with a width of $\Delta L = 0.01$\,pc along the vector defined in 1 above. 
Within each bin, the radial center is defined at the maximum density point. 
We then calculate the radial density profile within the small segment of the filament.
\item We examine if the line segments with gas mass $\Delta M_{\rm cell}$ are gravitationally unstable by comparing the local line mass, $M_{\rm line} = \Delta M_{\rm cell}/\Delta L$, with the critical line mass, $M_{\rm line,crit} = 2 c_{\rm s}^2/G$ \citep[e.g.][]{inutsuka92,inutsuka97}.
\end{enumerate}
Using the above procedures, we find gravitationally unstable line segments with gas masses $M_{\rm frag} = M_{\rm line} L_{\rm Jeans}$ = $170$--$410\,\msun$ (e.g., Figure~\ref{fig9} for CL3). 
These are identified as star-forming gas clouds.
We estimate the stellar mass formed in each fragment using a fitting formula derived from the results of a set of first star formation simulations \citep[eq.~17 in][]{hirano14}.
We assume the mean ratio of rotational energy to gravitational energy ($\beta_{\rm cloud} = 0.3$).\footnote{Although the fitting function is derived by using spherically averaged radial profiles, but we use the formula to estimate the stellar mass from the fragment mass.}
The estimated stellar masses, typically of several tens to a hundred solar masses, are listed in Table~\ref{tab2}.

\begin{figure*}[t!]
\begin{center}
\includegraphics[width=1\textwidth]{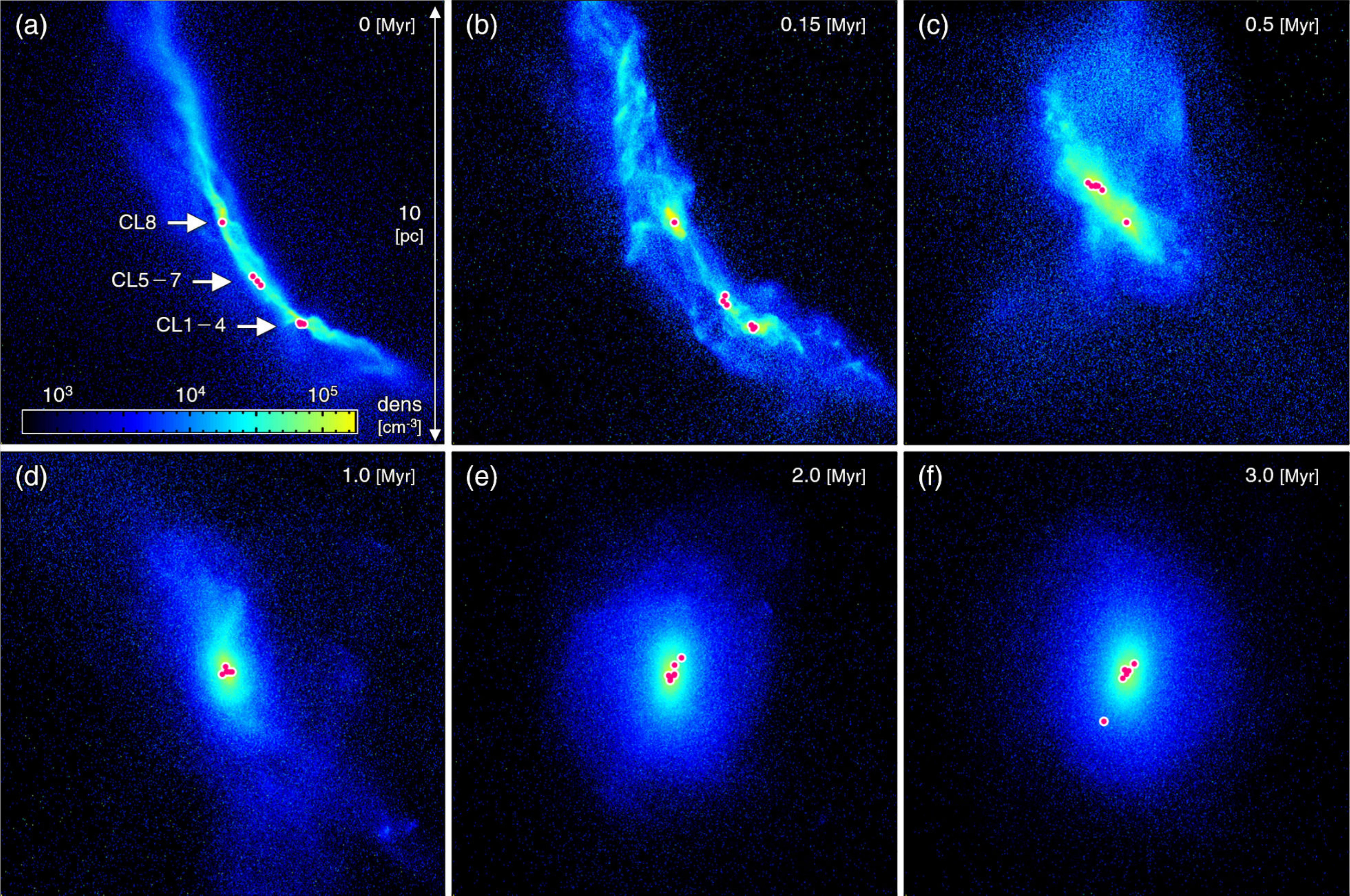}
\caption{
Time evolution of the filament structure during the $N$-body simulation at $t = 0$, $0.15$, $0.5$, $1$, $2$, and $3$\,Myr.
The color-scale shows the gas density distribution in a region of $10$\,pc on a side, and the dots represent the eight massive stars (CL1 to 8).
}
\label{f10}
\end{center}
\end{figure*}

We study the dynamical interaction of the massive stars using the hybrid $N$-body code {\sc bridge} \citep[][]{fujii07,sakurai17}.
The eight gas clumps (stars) are replaced with $N$-body particles with positions and velocities that are calculated from those of the original SPH particles.
The other SPH particles are converted to $N$-body particles with the same mass and velocity.
In the following $N$-body calculations, we consider neither gas drag nor pressure forces from the diffuse gas component within the host halo, which interacts with itself and the star particles only through gravity.
At the initial state, there are two multiple systems with four (CL1 to 4) and three (CL5 to 7) stars, and a single star (CL8; Figure~\ref{f10}(a)). 
Upon starting the $N$-body simulation, the eight stars gather quickly to form a cluster of massive primordial stars, and the whole filament collapses to form a spherical system (Figures~\ref{f10}(b) to (d)).

\begin{figure}[t!]
\begin{center}
\includegraphics[width=0.9\columnwidth]{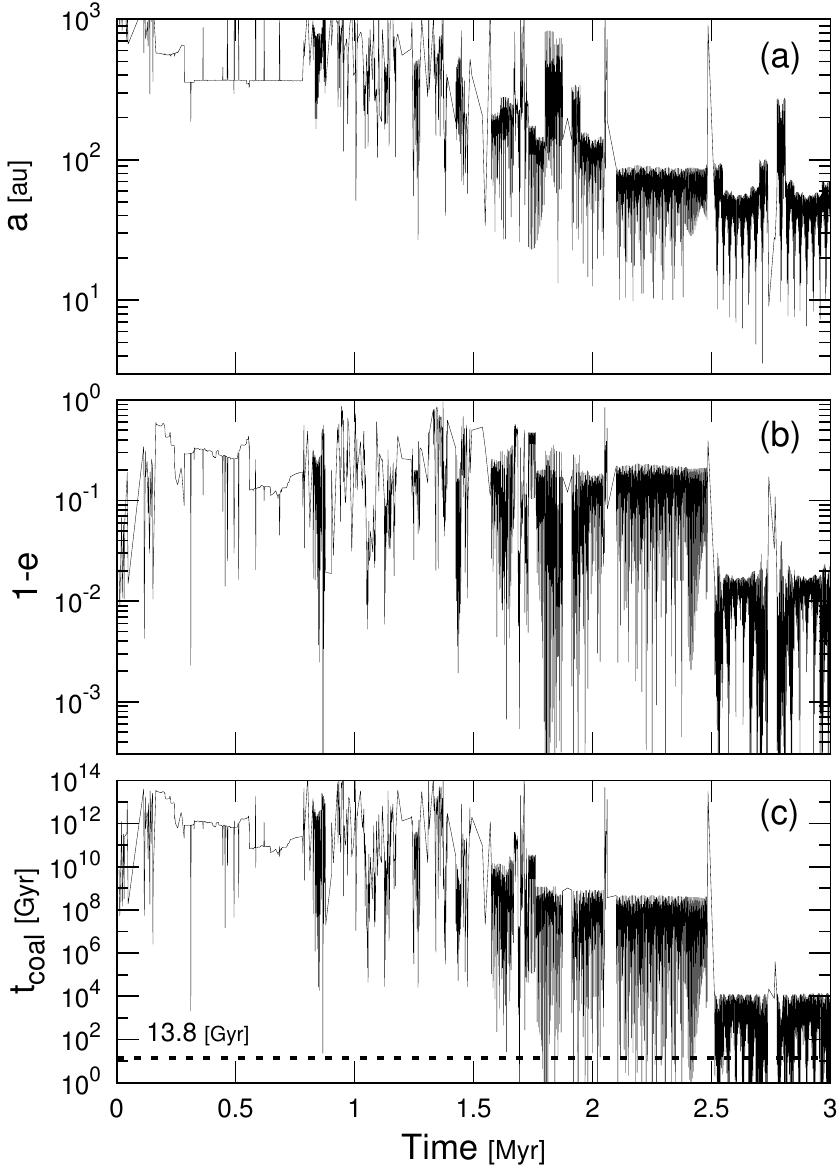}
\caption{
Time evolution of binary properties during the $N$-body simulation: semi-major axis (panel (a)), eccentricity (b), and timescale of the binary merger (equation~\ref{eq:t_coal_2}, panel (c)), respectively.
The dashed line in panel (c) represents the Hubble time, $13.8$\,Gyr.
}
\label{f11}
\end{center}
\end{figure}

An important process is the formation of hard binaries. 
Once a hard binary forms, multi-body interactions continuously occur and the binary becomes even harder. 
Figure \ref{f11} shows the time evolution of the semi-major radius and eccentricity of the hardest binary in the cluster. 
From $\sim\!1$ to $\sim\!2.5$\,Myr, the semi-major axis of the binary gradually decreases owing to repeated stellar encounters, through which the binary eccentricity increases.
The hardest binary interacts with the other stars, but the scattered stars remain bound within the host dark matter halo. 
After $\sim\!2.5$\,Myr, the binary evolution slows down, for the other stars are scattered out of the cluster. 
At $3$\,Myr, the hardest binary consisting of CL3 ($111\,\msun$) and CL6 ($86\,\msun$) has a semi-major axis of $56.1$\,au and eccentricity of $0.981$ (Figures~\ref{f10}(f) and \ref{f11}).

In order to see the effect of the assumed stellar mass, we have also performed $N$-body simulations with a fixed stellar mass. 
We set $m_{\rm star} = 30\,\msun$ in one case, and $m_{\rm star} = 100\,\msun$ in the other.
The two runs show a quite similar evolution of the system and binary formation.
We thus conclude that the hard binary formation is not a peculiar result of our $N$-body simulation with the estimated stellar masses. 
The precise stellar mass is not important in the star cluster evolution and in the formation of hard binaries.

The formed massive star binary can be a promising progenitor of BH-BH merger \citep[e.g. figure~2 in][]{belczynski17} like the recently detected GW sources \citep{GW170104}.
The merger timescale of a black hole binary depends on the separation \citep[equation~80 in][]{kinugawa14},
\begin{eqnarray}
  t_{\rm coal}(e_0 = 0) &=& 10~{\rm Gyr} \left( \frac{a_0}{0.2~{\rm au}} \right)^4 \notag \\
  && \left( \frac{M_1}{30~\msun} \frac{M_2}{30~\msun} \frac{M_1 + M_2}{60~\msun} \right)^{-1} \, ,
  \label{eq:t_coal}
\end{eqnarray}
where $e_0$ is the initial eccentricity, $a_0$ is the initial semi-major axis, and $M_1$ and $M_2$ are the masses of the primary and secondary stars. 
There is a strong constraint on the initial separation that allows coalescence within a Hubble time.
For dynamically formed binaries with large eccentricities, the coalescence time is given by
\begin{eqnarray}
  t_{\rm coal}(e_0) \sim (1-e_0)^{7/2} t_{\rm coal}(e_0 = 0) \, .
  \label{eq:t_coal_2}
\end{eqnarray}
Here we use equation~81 in \cite{kinugawa14}.
With $e_0 = 0.9$ and $0.99$, the necessary initial separation for merging within a Hubble time is $4.3$ and $31$ times the critical value for $e_0 = 0$.
Figure~\ref{f11}(c) shows the merger timescale calculated using equation~(\ref{eq:t_coal_2}).
Although the binary separation does not decrease significantly after $\sim\!2.5$\,Myr, the merger timescale periodically falls below the Hubble time because both the separation and eccentricity vary considerably.
In principle, one needs to follow the stellar evolution of the binary stars and more complex interaction with gaseous envelope in order to determine the final fate. 
Our $N$-body simulations do not provide accurate predictions for when exactly the binaries merge.  
However, the fact that eccentric close-binaries with separations less than $10$\,au are formed within a few million years suggests that a fraction of such massive star binaries likely leave massive BH binaries that can coalesce through emission of gravitational waves within a Hubble time.

\section{Discussions}
\label{sec:dis}

We study the formation of the first star clusters by performing cosmological hydrodynamic simulations.
Our simulations with early baryonic streaming motions show the formation of a massive, elongated filamentary cloud. 
Gravitational fragmentation of the large filament produces multiple gas clouds in which massive primordial stars of $\sim\!100\,\msun$ are formed.

The formation of the first star clusters offers an interesting possibility for direct observation. 
Although the system found in our simulation is small, with a total stellar mass of about a few thousand solar masses, if a larger filament is formed by a similar process, and if more massive and hence luminous star clusters are formed, they may be detected by future telescopes such as the James Webb Space Telescope \citep{tumlinson01,bromm11}, especially when the magnification effect by gravitational lensing is utilized \citep{rydberg13}.
Yet there is a more exciting implication for observations of the first star cluster systems.
Our simulations show the formation of a hard binary of massive stars at the center of the star cluster. 
If the remnant BHs with dozens of solar masses form an eccentric close-binary system, it emits strong gravitational waves with a characteristic signatures at the final merger phase \citep[][]{kinugawa14,kinugawa16b,inayoshi17}.
In fact, mergers of massive BHs at the present-day Universe have been already detected \citep{GW150914}, and the origin of the massive BHs could be the first stars.
Our results in the present paper provide a viable formation path for the massive BH binaries in a low- (zero) metallicity environment.
Intriguingly, \cite{GW170104} reported that the measured BH spins are likely misaligned. 
The binary system may have been formed through dynamical interactions in dense clusters, in a similar manner as studied in the present paper.

In our simulations, the baryonic streaming motions produce large filaments with lengths of a few tens of parsecs. 
Filamentary gas clouds can also be formed in different manners in variants of cosmological models.
For example, the formation of large filamentary structure and its fragmentation are found in simulations of warm dark matter cosmology \citep{gao07wdm,gao15} and also in a model with axion-like particle dark matter \citep{hirano17fdm}.
It would be interesting to follow the dynamics of the fragments and possible formation of star clusters in these cosmological models.

Vigorous fragmentation of a primordial gas cloud can occur in a different environment. 
It is thought that an initially ``hot'' gas cloud with a temperature of $\sim\!8,000$--$10,000$\,K can be formed in an atomic-cooling halo under the influence of an external Lyman-Werner radiation \citep{omukai01fuv}.
If the external radiation intensity is below a certain critical value for the complete suppression of hydrogen molecule formation, the thermal evolution of a gas cloud is similar to those found in our simulations \citep{latif15}.
The combined effects of the baryonic streaming motions and (weak) ultraviolet radiation may well produce the conditions for primordial star cluster formation.
Cosmological simulations with realistic initial conditions with the baryonic streaming motions provide the opportunity to explore a variety of formation paths of the first stars, star clusters, and galaxies.

\acknowledgments 
We would like to thank Oliver Hahn for discussion about {\sc music} and Kohei Inayoshi and Tomoya Kinugawa for stimulating discussions.
The numerical calculations were carried out on Cray XC30 at Center for Computational Astrophysics (CfCA), National Astronomical Observatory of Japan and COMA at Center for Computational Sciences, University of Tsukuba. 
This work was financially supported by JSPS KAKENHI grants 14J02779 to S.H.; 15J08816 to Y.S.; 26800108 and 17H06360 to M.S.F.; by JST CREST JPMHCR1414, MEXT Priority Issue 9 on Post-K Computer to N.Y.; and by Advanced Leading Graduate Course for Photon Science to Y.S..

\bibliographystyle{aasjournal}
\bibliography{biblio}

\end{document}